\documentclass[preprint]{aastex}
\usepackage{amsmath}
\usepackage{amssymb}
\usepackage{graphicx}
\usepackage{subfigure}
\usepackage{epsfig}
\usepackage{color}
\usepackage{fullpage}

\begin{document}

\title{Physical and Morphological Properties of [O~II] Emitting Galaxies in the HETDEX Pilot Survey}

\author{Joanna Bridge\altaffilmark{1}, Caryl Gronwall\altaffilmark{1}, Robin Ciardullo\altaffilmark{1}, Alex Hagen\altaffilmark{1}, Greg Zeimann\altaffilmark{1}, A. I. Malz\altaffilmark{1}}
\affil{Department of Astronomy \& Astrophysics, The Pennsylvania State University, University
Park, PA 16802}
\email{jsbridge@psu.edu, caryl@astro.psu.edu, rbc@astro.psu.edu, hagen@psu.edu, grzeimann@psu.edu, aimalz@psu.edu}
\author{Viviana Acquaviva}
\affil{Department of Physics, New York City College of Technology, City University of New York,
Brooklyn, NY 11201}
\email{vacquaviva@citytech.cuny.edu}
\author{Donald P. Schneider\altaffilmark{1}}
\affil{Department of Astronomy \& Astrophysics, The Pennsylvania State University, University
Park, PA 16802}
\email{dps@astro.psu.edu}
\author{Niv Drory, Karl Gebhardt, Shardha Jogee}
\affil{Department of Astronomy, University of Texas at Austin, Austin, TX 78712}
\email{drory@astro.as.utexas.edu, gebhardt@astro.as.utexas.edu, sj@astro.as.utexas.edu}

\altaffiltext{1}{Institute for Gravitation and the Cosmos, The Pennsylvania State University, University Park, PA 16802}

\begin{abstract}
The Hobby-Eberly Dark Energy Experiment pilot survey identified 284 [O~II] $\lambda$3727 emitting galaxies in a 169 arcmin$^2$ field of sky in the redshift range $0 < z < 0.57$. This line flux limited sample provides a bridge between studies in the local universe and higher-redshift [O~II] surveys. We present an analysis of the star formation rates (SFRs) of these galaxies as a function of stellar mass as determined via spectral energy distribution fitting. The [O~II] emitters fall on the ``main sequence" of star-forming galaxies with SFR decreasing at lower masses and redshifts. However, the slope of our relation is flatter than that found for most other samples, a result of the metallicity dependence of the [O~II] star formation rate indicator. The mass specific SFR is higher for lower mass objects, supporting the idea that massive galaxies formed more quickly and efficiently than their lower mass counterparts.  This is confirmed by the fact that the equivalent widths of the [O~II] emission lines trend smaller with larger stellar mass. Examination of the morphologies of the [O~II] emitters reveals that their star formation is not a result of mergers, and the galaxies' half-light radii do not indicate evolution of physical sizes. 
\end{abstract}


\section{Introduction}
Numerous surveys have shown that the star formation rates (SFRs) of galaxies evolve with redshift, from $z\sim6$ \citep[e.g.,][]{ham97, hop04, gonz10, bouw11} down to the local universe.  From $z\sim2$ to $z=0$, the average stellar mass of galaxies that are actively star-forming, along with their SFRs, have decreased steadily \citep[e.g.,][]{sea73, cow96, heav04, noes07a, lop10, whit12, pirz13}.  While considerable effort has been expended on studying star-forming galaxies either in the local universe or above $z\sim1$, relatively few surveys have focused on the $\sim$ 7 Gyr in between the two epochs \citep[e.g.,][]{wolf05, jog09, rob10, lotz11}.

One star formation rate indicator that is particularly useful in this range is the doublet [O~II] emission at $\lambda 3727$  \citep[e.g.,][]{ken98, kew04, ke12}, which is collisionally excited by the ionized electrons of H~II regions.  At $z \lesssim 1$, [O~II] is easier to detect than the rest-frame UV, while being less sensitive to the effects of time-averaging of the star-formation rate. It is, however, less straightforward to use, as its strength can be dependent on metallicity and the ionization parameter.

As part of the pilot survey for the Hobby Eberly Telescope Dark Energy Experiment (HETDEX), \citet{ad11} identified 397 emission-line galaxies over 169 arcmin$^2$ of sky.  While the focus of this pathfinding study was the identification of Ly$\alpha$ emitters (LAEs) at $1.9 < z < 3.6$ \citep{hill08, blan11}, the observations also detected a large number of lower-redshift [O~II] emitting galaxies.  This resulted in spectra for a well-defined, emission-line flux limited sample of 284 [O~II] emitters between $0 < z < 0.57$.  While a number of star formation rate indicators have been previously used in this redshift range \citep{ken98, pet01, hop04}, this samples provides a unique resource for connecting low- and high$-z$ observations with a single, consistent SFR indicator.  \cite{ciar13} has examined the [O~II]-based star-formation rates for these galaxies and concluded that, although the scatter between the UV and [O~II] SFR estimators is significant ($\sigma \sim 0.3$), there is no systematic offset between the techniques.

The HETDEX survey will shortly be yielding data for an emission-line selected sample of $\sim 10^6$ [O~II] emitting galaxies in the redshift range $0 < z < 0.5$, with wavelength coverage from 3500~\AA\ to 5500~\AA\null.  In this paper, we present a pathfinding analysis of the physical and morphological properties of the [O~II] emitting galaxies in the HETDEX Pilot Survey (HPS\null).   In \S 2, we summarize this survey, and the techniques used to identify the 284 [O~II] line emitters in the sample. In \S 3, we model the spectral energy distributions (SEDs) of these galaxies to produce estimates of their stellar mass, internal reddening, and star-formation rate.  In \S 4, we discuss the evolution of the star formation rates in the last $\sim 5$~Gyr of cosmic time, and how the relationship between stellar mass and star-formation rate changes over our survey epoch.  In \S 5, we compare the [O~II] equivalent widths of our galaxies to their stellar mass, and show that the star-forming main sequence is reflected via an anti-correlation between these two parameters.  In \S 6 and 7, we discuss the morphology and size distribution of the HPS [O~II] galaxies.  We show that out to $z \lesssim 0.5$, there is no evidence for significant size evolution or merger-driven starbursts.

Throughout this work, a flat $\Lambda$CDM cosmology is used with $\Omega_m =  0.3$, $\Omega_\Lambda = 0.7$, and $H_0 = 70$ km s$^{-1}$ Mpc$^{-1}$ \citep{kom11}.

\section{The HETDEX Pilot Survey Data}
The data used here were taken as part of the HETDEX pilot survey (HPS; \citealt{ad11}).  The survey focused on four different regions of the sky, coincident with the fields of the Cosmological Evolution Survey (COSMOS; \citealt{scov07}), the Great Observatories Origins Deep Survey North (GOODS-N; \citealt{gia04}), the Munich Near-IR Cluster Survey (MUNICS; \citealt{dror01}), and the XMM Large-Scale Structure survey (XMM-LSS; \citealt{pier04}). HPS utilized the fiber-fed George and Cynthia Mitchell Spectrograph mounted on the Harlan J. Smith 2.7-m telescope at the McDonald Observatory.  This spectrograph is a prototype for the Visible Integral-field Replicable Unit Spectrograph (VIRUS) that will eventually be used for the full HETDEX survey \citep{hill08}. The instrument contains 246 4\farcs2-diameter fibers, which produced spectra from 3500 to 5800~\AA~at a 5-\AA~full-width-half-maximum (FWHM) spectral resolution. For full survey details, see \cite{ad11} and references therein.

HPS was a blind integral field spectroscopic survey that searched for objects with emission lines within each exposure. When an emission line was found, its optical counterpart was identified in the ancillary images of COSMOS, GOODS-N, XMM-LSS, and MUNICS using a Bayesian matching scheme.  As confirmation of the validity of the counterpart identification, for the 200 objects in the GOODS-N and COSMOS fields, we compared the redshifts obtained from the HPS spectra to spectroscopic or photometric redshifts of the counterparts obtained from the literature.  In almost all cases there was complete agreement, thus confirming the identification process. Equivalent widths  were estimated by comparing the HPS line flux to the flux density of the photometric continuum.  The emission lines and redshifts were then identified using either other lines in the spectrum, or, since most objects possessed only a single line, an equivalent width cut.  If only a single line is present in the spectrum, it is by necessity either [O II] or Ly$\alpha$ because given the spectral range of HPS, only those two emission lines would present as a single line \cite[]{ad11}.  For example, if the emission line was [O III], other lines such as H$\beta$ and probably [O II] would also be present.

Objects with only a single emission line detection could be either an [O II] galaxy with $z < 0.57$ or an LAE with $1.9 < z < 3.8$, as described in Adams et al. (2011).  We apply an equivalent width (EW) cut of a rest-frame EW $>$ 20~\AA~to identify and reject a higher redshift range LAE. A rest-frame EW $> 20$~\AA~is typical for narrow-band surveys of LAEs \citep[e.g.,][]{gron07, ci12}. This rest-frame EW translates to an observed-frame EW $>$ 58 to 96~\AA~for $1.9 < z < 3.8$ LAEs. To confirm that our remaining galaxies are indeed lower redshift [O II] emitters and not higher redshift Ly$\alpha$ interlopers, we applied the objects' spectral energy distribution: high-redshift LAEs have no flux in the ultraviolet, as this light is shortward of the Lyman break at 912~\AA~in the rest frame.

We are confident that the observed-frame 58~\AA~EW cut to reject LAEs does not remove a significant high-EW tail of our [O~II] emitters: Hogg et al. (1998) found that $\sim$ 2\% of [O II] emitters have EW$_{rest} > 60$~\AA; our sample has a median redshift of $z=0.37$, where a 58~\AA~observed-frame EW translates to EW$_{rest}$ of 42~\AA.  Considering that the [O~II] distribution of EW e-folds at 20~\AA~for the median redshift of the sample \cite{ciar13}, there will be very few, if any, [O~II] emitters that are close to the Hogg et al.\ limit.

A total of 284 [O~II] emitting galaxies were identified with redshifts between $0.078 < z < 0.563$. Ninety percent  of the pointings of the survey reached a monochromatic flux limit of $1.0\times10^{-16}$ erg cm$^2$ s$^{-1}$, allowing us to recover 95\% of all objects with observer's-frame equivalent widths greater than 5~\AA~\citep{ad11, ciar13}. The survey volume probed for the [O~II] emitters is $4.24\times10^4$ Mpc$^3$ \cite[]{ad11}.

Thirty of the 284 [O~II] emitters ($\sim10$\%) have X-ray counterparts from surveys by \emph{Chandra} and \emph{XMM-Newton} \cite[]{ad11}. This is not surprising considering the depth of the X-ray observations, particularly in the GOODS-N field, where many of the X-ray emitters were located.  \cite{ciar13} showed that for most of these sources, the X-rays are likely associated with normal star formation, but 10 ($\sim3$\% of the sample) have X-ray luminosities greater than $10^{40}$ erg s$^{-1}$ in the 2 to 8 keV band. These are likely AGN; subsequent analysis shows that while these objects are not necessarily the brightest or the most massive in this sample, they do exhibit anomalously greater [O II]  star formation rates, especially at the higher redshifts in the sample. They have therefore been excluded from our analysis. For further discussion, see \cite{ciar13} and sources therein.

\section{Spectral Energy Distribution Fitting}
Physical properties of the [O~II] emitters were determined by fitting the SEDs of the galaxies using the Markov Chain Monte Carlo (MCMC) code GalMC \cite[]{acq11}. Standard $\chi^2$ minimization works best if the probability distribution of the parameters is a Gaussian, and there are few variables to be fit, as the computation time grows approximately exponentially with the number of parameters \cite[]{ser11}.  Additionally, $\chi^2$ minimization will fail to identify double-valued solutions. MCMC is advantageous because it makes no assumption about the underlying probability distribution, and any bimodal solutions are immediately obvious because in such a case, the fit will fail to converge. See \cite{acq11} for a detailed explanation of GalMC and the MCMC algorithm.

The SED fitting was performed using up to 11 photometric bands from various sources covering a range of wavelengths from the far-UV to the IR\null.  Table~\ref{tbl-1} lists the data that were used to perform the fits for the four survey regions.  Most of the photometry was taken from \cite{ad11}.  The UV bands were found in the GR6 catalog of \emph{GALEX} \cite[]{mart05}, while the data from the \emph{Spitzer} 3.6 $\mu$m and 4.5 $\mu$m bands were gathered from publicly available catalogues.  For COSMOS, this was the S-COSMOS survey \cite[]{sand07}, for XMM-LSS it was the \emph{Spitzer} Wide-area Infra-Red Extragalactic survey (SWIRE; Lonsdale et al.\ 2003), and for GOODS-N, the IR was part of the original survey \cite[]{gia04}. Aperture and PSF effects were accounted for in the photometry by the creators of the catalogs. 

The GalMC SED fitting code uses the stellar population models of \cite{bc03} that were updated in 2007 to incorporate improved models of thermally pulsing stars on the asymptotic giant branch (TP-AGB) (Charlot \& Bruzual 2010, private communication). A constant star formation history (SFH) was assumed, along with a Salpeter (1955) initial mass function (IMF) (with M$_{\text L}$ = 0.1 M$_{\sun}$ and M$_{\text U}$~=~100~M$_{\sun}$), and a Calzetti extinction law \cite[]{calz00}.  Nebular emission, both in the continuum and for lines, was included in proportion to the rate of H-ionizing photons, with the relative line intensities of H, He, C, N, O and S being a function of metallicity (see Acquaviva et al.\  2011 for further detail). The metallicity was fixed at solar. The \cite{lew02} GetDist program from the publicly available CosmoMC software was used to analyze the chains output by GalMC and test for convergence via the \cite{gel92} $R$ statistic.  The criterion employed here is $R-1 < 0.2$, as prescribed by \cite{acq11}.  Additionally, seven objects with fewer than five photometric bands were excluded from the sample in order to ensure robust SED fits. The parameters obtained from the SED fitting were stellar mass, age since the onset of star formation, and reddening.  

It is important to consider how much the results of the SED fits depend on the input models and parameters.  \cite{con13} states that the stellar mass is the most robust parameter that can be found with SED fitting techniques.  He notes that generally, stellar mass can vary by $\sim0.3$ dex for star-forming galaxies, depending on the SFH chosen, and by as much as 0.6 dex in extreme cases. \cite{acq11}, however, state that when testing the robustness of GalMC, the choice of a constant star formation history does not result in significant differences from that when using exponentially increasing or decreasing SFHs.  Additionally, \cite{pirz12} found that the choice of IMF and stellar population model using the MCMC SED fitting algorithm $\pi$MC$^2$ resulted in parameter variation of a factor of a few.  Metallicity estimates from SED fitting are uncertain at best, as stars of varying metallicities contribute to the overall metallicity of the galaxy in different ways as different times \cite{acq11}.  Fixing the metallicity at solar can affect the derived ages by up to 0.5 dex, but, more importantly, affects the stellar mass by only $\pm0.1$ dex \cite{wu09}. This falls easily within the mass error estimates. Therefore, while the analysis presented here may contain some systematic differences, the trends will remain the same. 

\begin{figure}[t!]
\centering
\scalebox{0.7}
{\epsfig{file = 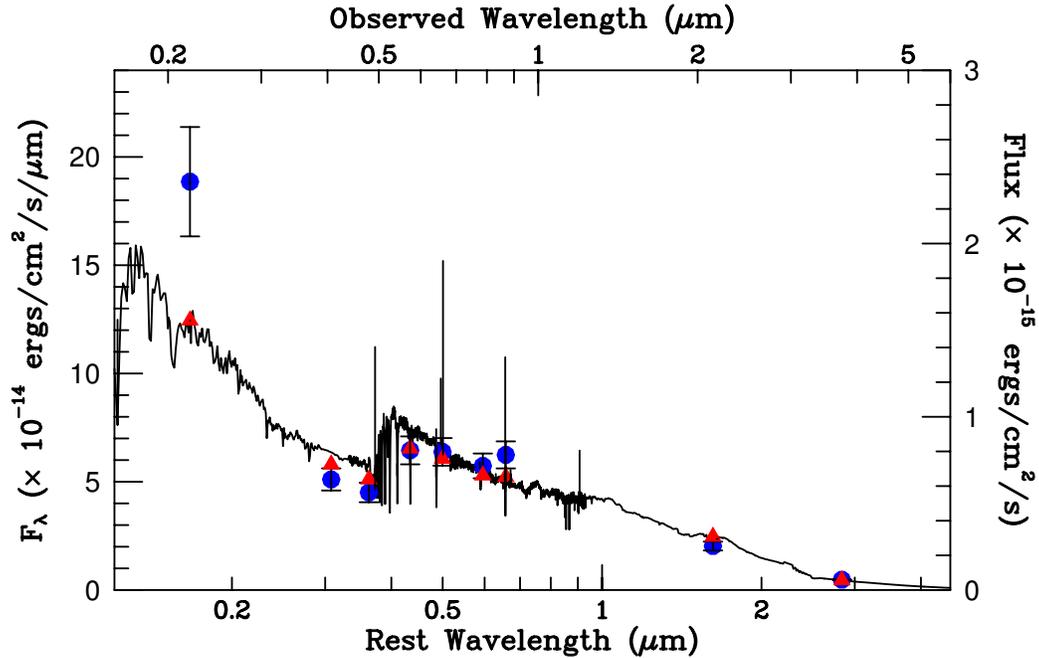}}
\caption{A typical SED fit plotted with the photometry of an [O~II] emitter ($z = 0.32$) from the COSMOS field.  The blue circles are the observed flux densities, the red triangles are the fitted SED flux densities, and the black line is the SED fit.}
\label{fig:phot}
\end{figure}

\begin{figure}[p]
\centering
\scalebox{0.7}
{\epsfig{file = 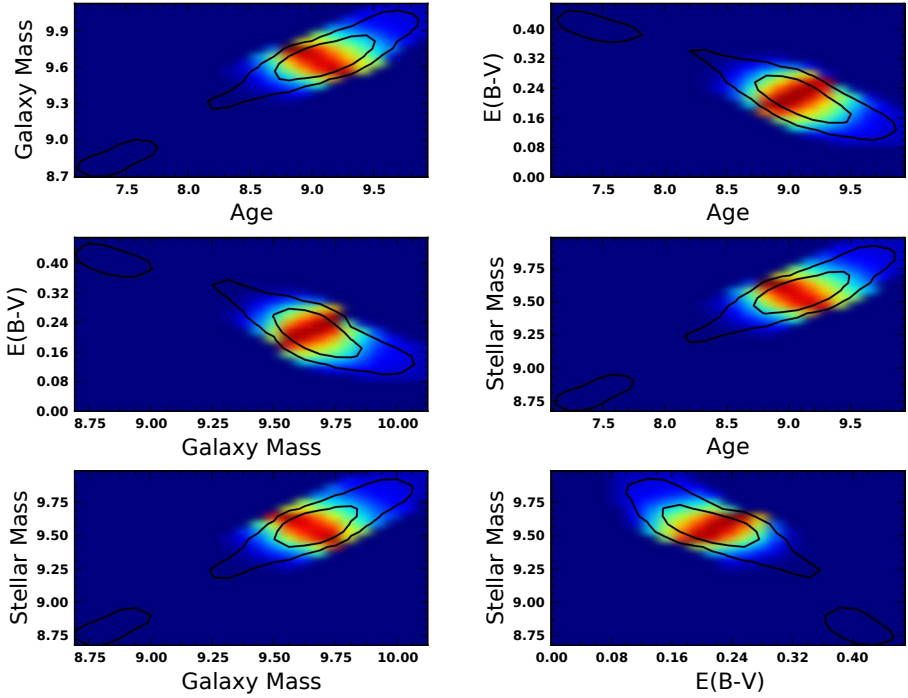}}
\caption{Two-dimensional confidence levels for the parameters obtained from the SED fit for the same galaxy as in Figure~\ref{fig:phot}.  The black contours indicate the 68\% and 95\% confidence regions, while the color gradient is based on average likelihood. The colors follow the likelihood of the two parameters plotted, while the black contours indicate the probability distribution weighted by the parameters not represented in the plot. For flat priors, lack of exact overlap indicates that the posterior distribution is non-Gaussian; in this case, the contours also show a bi-modal probability distribution. The axes of these ellipse-like curves indicate degeneracies. Galaxy mass is defined as the integral of the instantaneous star formation over the lifetime of the galaxy, while stellar mass takes into account mass loss and stellar lifecycles, and is the parameter used throughout this analysis.}
\label{fig:sed}
\end{figure}

\begin{figure}[h]
\centering
\scalebox{0.8}
{\epsfig{file = 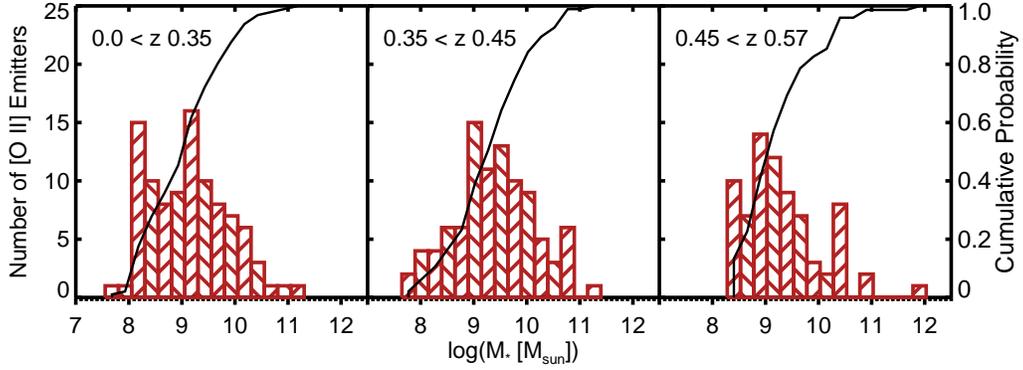}}
\caption{Distribution of the stellar masses of the HPS [O~II] emitters in three redshift bins.  The black line is the cumulative probability distribution. See Table~\ref{tbl-2} for the median masses in each redshift bin.}
\label{fig:mass}
\end{figure}

The results of the mass SED fitting are given in Table~\ref{tbl-3}. A sample fit for a COSMOS [O~II] emitter ($z = 0.32$) is given in Figure~\ref{fig:phot}.  Figure~\ref{fig:sed} shows the two-dimensional confidence contours of the various parameters output by the SED fitting algorithm for the same galaxy. For the purposes of this work, the stellar mass ($M_*$) is the important parameter.  The stellar mass distribution of the HPS [O~II] emitters is shown in Figure~\ref{fig:mass}, where the masses have been divided into redshift bins.  As indicated by the cumulative probability distributions, the mass distribution is fairly consistent across the entire redshift range, demonstrating the ability of an emission-line selected survey to detect low mass galaxies at higher redshifts.  The median masses of the galaxies divided by redshift bin are given in Table~\ref{tbl-2}.

\section{Internal Extinction}
Accurate determination of the the internal extinction in each galaxy is necessary to get reliable measurements of its [O~II]-based star formation rate .  Ideally, the extinction can be calculated directly from the Balmer decrement.  Unfortunately, the HPS spectral range does not include H$\alpha$, and the wavelength baseline between H$\beta$ and the other (weaker) Balmer lines is insufficient for our purpose.  We therefore must rely on stellar-based reddening estimates; specifically, the rest-frame UV continuum slope $\beta$. Between 1250~\AA\ and $\sim 2800$~\AA, the intrinsic slope of a star-forming galaxyÕs continuum can be well-fit via a power law where $f_{\lambda} \propto \lambda^{\beta}$.  Using the \emph{GALEX} FUV and NUV bands, and in some cases the $u^*$ or $U$ (depending on the field) we calculated $\beta$ for 215 galaxies in the sample.  Any flattening of this relation is most likely due to extinction, with the relationship between $\beta$ and dust extinction \cite[]{mer99} given by
\begin{equation}
A_{1600} = 4.43+1.99\beta
\end{equation}
This assumes an intrinsic UV spectral slope of $\beta_0$ = -2.23, consistent with a constantly star-forming population. The reddening is then simply
\begin{equation}
E_s(B-V) = A_{1600}/k_{1600}
\end{equation}
where $E_s(B-V)$ is the color excess of the stellar continuum and $k_{1600} = 9.97$ is the reddening curve given by \cite{calz00}.

The other 69 galaxies lacked sufficient photometry in the UV for their redshifts to do a robust UV slope calculation.  In this case, we employed the mass-extinction relationship of \cite{garn10}. Using Data Release 7 of the Sloan Digital Sky Survey (SDSS), they modeled the dependency of extinction on mass as
\begin{equation}
A_{H\alpha} = 0.91 + 0.77X+0.11X^2-0.09X^3
\end{equation}
where $X =$ log($M_*/10^{10}$ M$_\odot$). Note that since this relation is a cubic, it is only valid over a specific stellar mass range; for the thirteen galaxies below $10^{8.5}$ M$_\odot$ for which we do not have UV slope measurements, we assume an $E(B-V)$ set to the minimum reddening defined by the relation in Equation 3. We then employ Equation 2 to determine $E(B-V)$ using the appropriate reddening curve.  This relation between stellar mass and dust has been shown to be valid out to $z \sim1.6$ \cite[]{zahid13}.

Finally, the entire sample was corrected for extinction using the \cite{calz00} extinction law 
\begin{equation}
F_i(\lambda) = F_o(\lambda)10^{0.4E_s(B-V)k(\lambda)}
\end{equation}
Note that $E_s(B-V)$ is related to the color excess of the nebular emission lines, such as [O~II], by
\begin{equation}
E_s(B-V) = 0.44E(B-V)
\end{equation}
\cite[]{calz00}

\section{Star Formation Rates}
There has been extensive discussion in the literature about the ``main sequence" of star-forming galaxies, i.e., the correlation between a galaxy's SFR and its stellar mass \citep[e.g.,][]{noes07a}. Both the slope and, more strongly, the normalization of the sequence has also been shown to evolve with redshift \cite[]{whit12}.

The SFRs for the HPS [O~II] emitters were calculated using the [O~II] $\lambda3727$ \cite{kew04} relation 
\begin{equation}
\textrm{SFR([O~II])}(\textrm{M}_{\sun}\:\textrm{yr}^{-1}) = (6.53 \pm 1.65)\times10^{-42} L(\textrm{[O~II]})\:(\textrm{ergs s}^{-1})
\end{equation}
where $L(\textrm{[O~II]})$ is the galaxy's [O~II] luminosity. Note that this is a mean calibration since the wavelength range for HPS does not include the H$\beta$ or [O III] $\lambda5007$ emission lines that would allow fine-tuning for the effects of variations in oxygen abundance and ionization states. 

\begin{figure}[!htbp]
\centering
\scalebox{0.8}
{\epsfig{file = 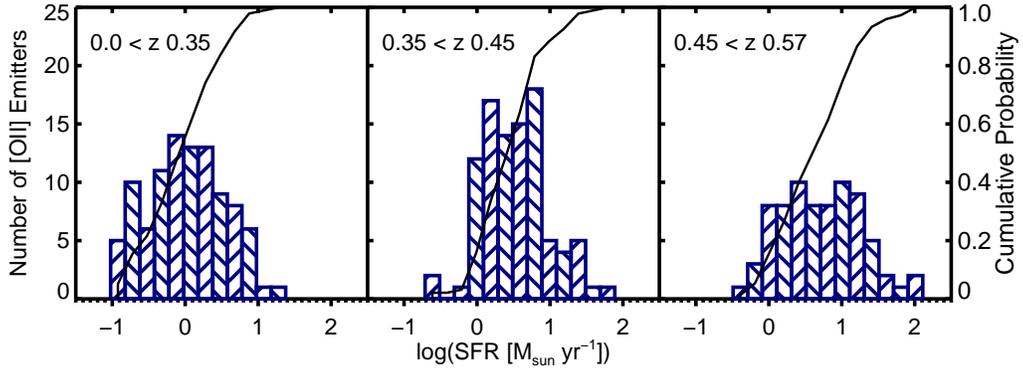}}
\caption{Distribution of the HPS [O~II] SFRs calculated via their [O~II] line luminosities in three redshift bins. The black line is the cumulative probability distribution.  See Table~\ref{tbl-2} for the median SFRs in each redshift bin. The shifting distribution shows a clear evolution of the high star formation end of this function.}
\label{fig:SFR}
\end{figure}

\begin{figure}[!htbp]
\centering
\scalebox{0.6}
{\epsfig{file = 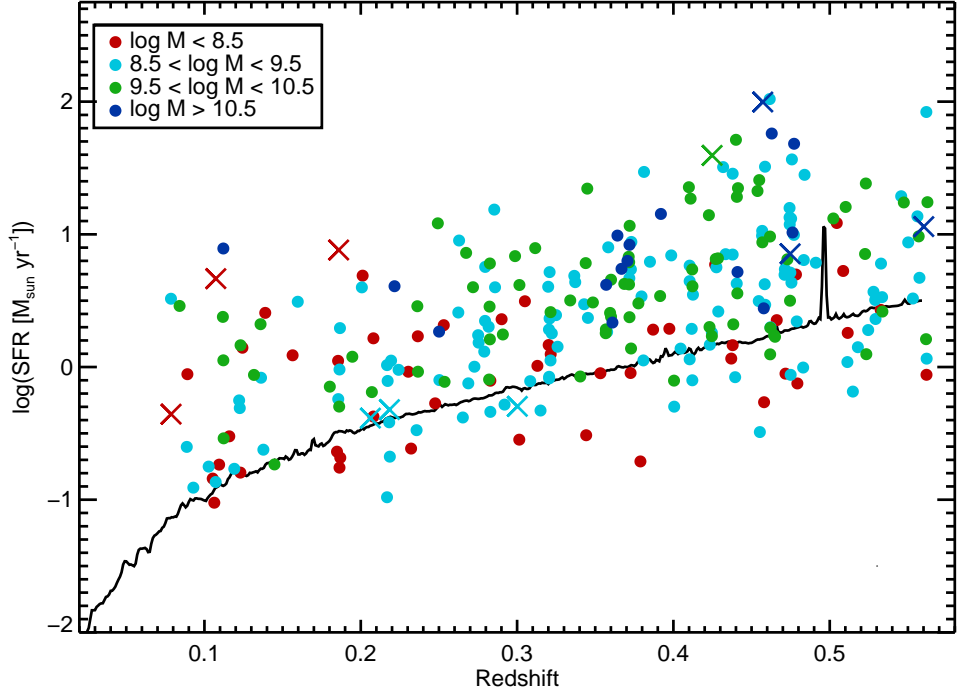}}
\caption{Evolution of the SFRs of the HPS [O~II] emitters as a function of redshift, color coded by mass. The black line represents the  detection limit as a function of redshift, where 80\% of the survey frames have 5$\sigma$ detection limits brighter than this threshold. The prominent [O I] airglow feature at 5577 \AA~at $z\sim0.5$ is clearly visible. The crosses are possible AGN candidates due to their bright X-ray emission. The representative SFR  error bar is shown.}
\label{fig:SFRvz}
\end{figure}

\begin{figure}[!htbp]
\centering
\scalebox{0.6}
{\epsfig{file = 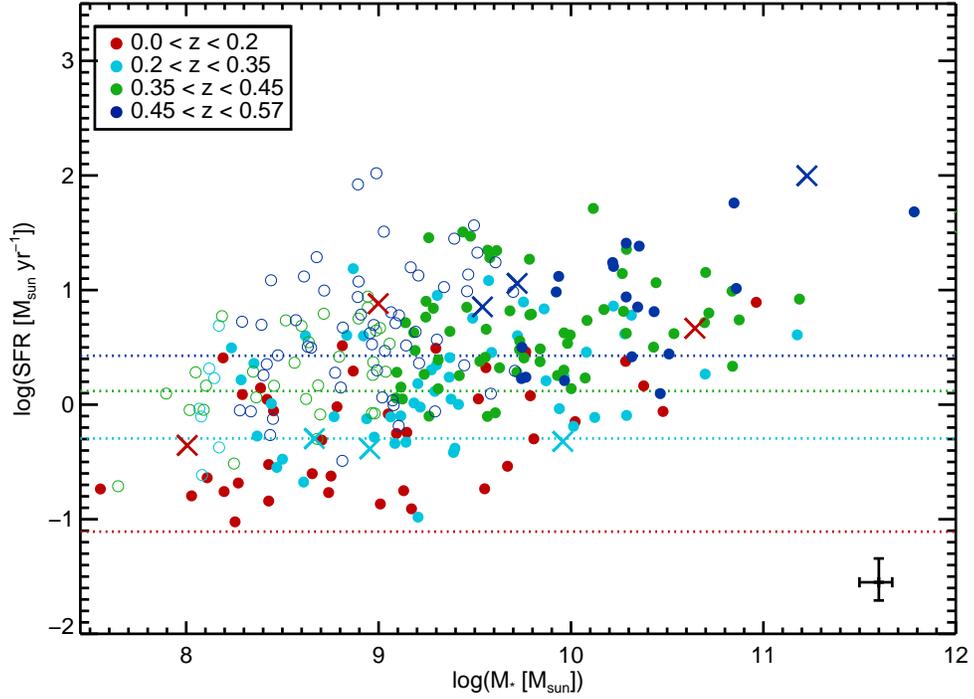}}
\caption{SFR vs. stellar mass for the HPS [O~II] emitters. The crosses are possible AGN candidates due to their bright X-ray emission. Representative error bars are shown. The open circles represent galaxies below the stellar masses where SFR become incomplete in each redshift bin.}
\label{fig:SFRvM}
\end{figure}

The distribution of the SFRs of the galaxies in our sample is given in Figure~\ref{fig:SFR}. The median SFRs of the [O~II] emitters are given in Table~\ref{tbl-2}. The evolution of star formation with redshift for the [O~II] emitters is presented in Figure~\ref{fig:SFRvz}, partially reflecting the changing luminosity limit of the survey.  However, after taking this into account, there is a clear relationship between the SFR of the galaxies and their redshifts, with the star formation tending to decrease at lower redshifts. Figure~\ref{fig:SFRvM} compares the SFR and mass of the HPS [O~II] emitters.  

\begin{figure}[!htbp]
\centering
\scalebox{0.6}
{\epsfig{file = 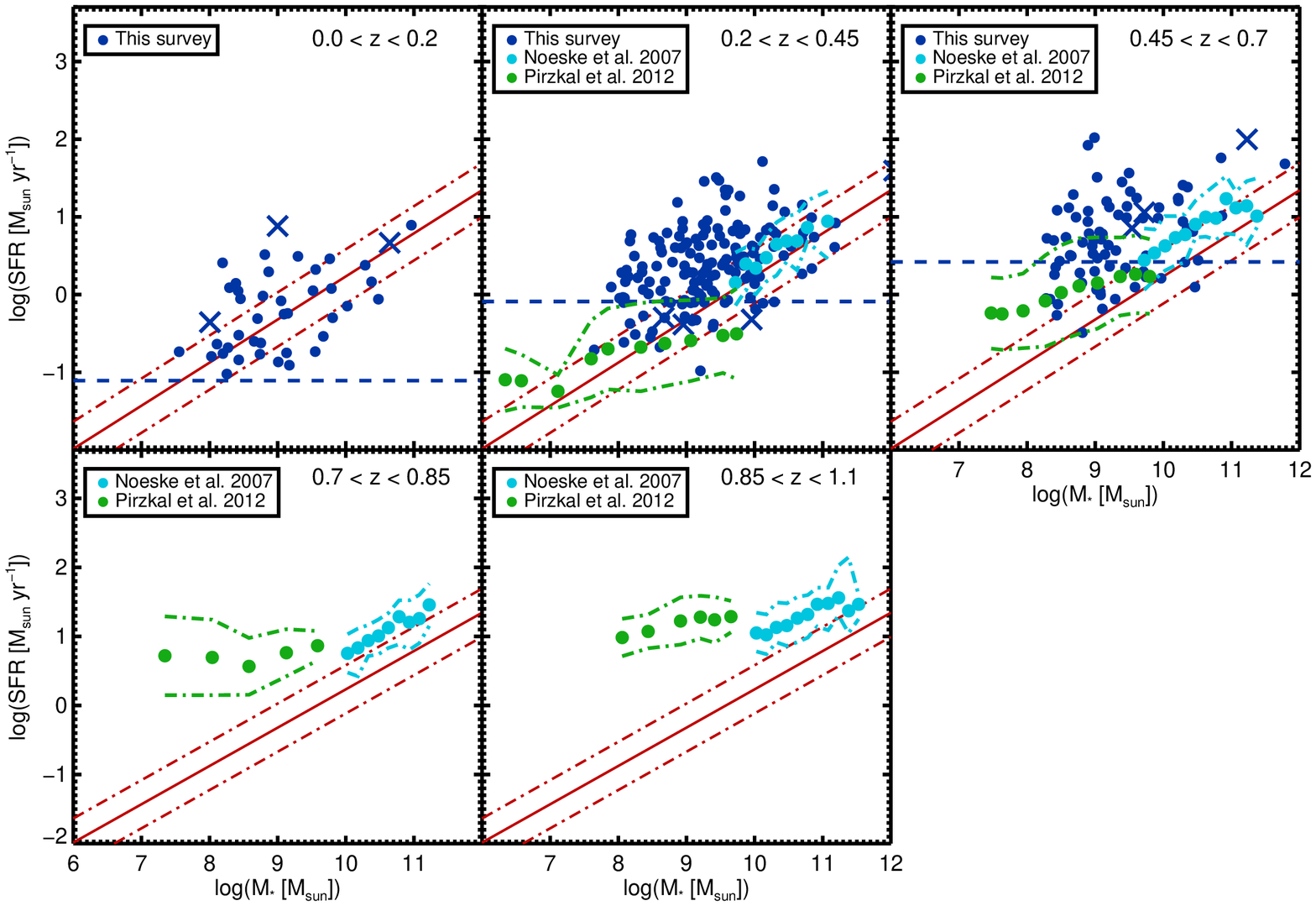}}
\caption{SFR vs. stellar mass for the HPS [O~II] emitters, as well as the \cite{noes07a} and \cite{pirz13} median values. The darker blue circles represent the HPS galaxies and the dark blue crosses are the possible AGN in the sample.  The green circles are the \citeauthor{pirz13} median values, and the light blue circles are the \citeauthor{noes07a} median values.  The light blue and green dash-dotted lines show $\pm1\sigma$ of the median values shown. The darker blue dashed line represents the 80\% completeness limit for the middle redshift of each plot. The red solid line is taken from \cite{lop13} and denotes the $M_*$-SFR relation for local galaxies up to $z\sim0.1$, and the red dash-dotted lines are $\pm 1\sigma$ of that relation. This relation is included in the higher redshift bins only to guide the eye.} 
\label{fig:SFRvM_PN}
\end{figure}

Previously, \cite{noes07a} and \cite{pirz13} have examined the main sequence of star-forming galaxies at slightly higher redshifts.  \cite{noes07a} analyzed a magnitude-limited set of star-forming galaxies from the Extended Groth Strip (AEGIS), probing galaxies of high mass within a redshift range of $0.2 < z < 1.1$ using \emph{Spitzer} (MIPS) 24 $\mu$m imaging as well as H$\alpha$, H$\beta$, and [O~II] emission lines. Similarly, \cite{pirz13} studied emission-line galaxies (ELGs) from the GOODS-N and -S fields as part of the Probing Evolution And Reionization Spectroscopically (PEARS) survey.  This slitless grism survey identified H$\alpha$, [O III], and [O~II] in the redshift ranges of $0 < z < 0.5$, $0.1 < z < 0.9$, and $0.5 < z < 1.5$, respectively.  It should be noted that in their determination of stellar mass via SED fitting, Pirzkal et al.\ assumed a constant SFH, as was assumed in this survey, while \citeauthor{noes07a} used an exponentially decreasing SFH. Figure~\ref{fig:SFRvM_PN} compares these data to our own measurements to show the progression of the star formation sequence. The addition of the HPS [O~II] emitters extends the galaxy star-forming main sequence of galaxies down to low redshifts, allowing for comparison of SFRs over a broad range of masses and redshifts.  As with the other galaxies, the [O~II] emitters show an evolution of SFR with redshift that is consistent with a galactic SFR evolution along the main sequence of star-forming galaxies.

To determine the intrinsic slope and scatter of the main sequence, we must make mass cuts to account for incompleteness in the star formation rates. This is an expected characteristic of an emission-line selected sample: while the underlying distribution may have a given slope, if a sample is limited by the SFRs of the galaxies, the apparent shape of the SFR-M${_*}$ relation will be skewed flatter than that for an unbiased sample.   Conversely, a mass-selected sample, such as the AEGIS survey, will produce a steeper slope.  To address this, we performed mass cuts in the four redshift bins represented in Figure~\ref{fig:SFRvM} by implementing an iterative algorithm using the slope and scatter of distribution to determine where the cuts should be made.  For $z < 0.2$, we made no cuts, as we expect the data are essentially complete at this low redshift.  For $0.2<z<0.35$, we invoked a mass cut at $10^{8.2}$ M$_\odot$, for $0.35 < z<0.45$ we used $10^{9.1}$ M$_\odot$, and for the highest redshift bin of $0.45<z< 0.57$ we used $10^{9.7}$ M$_\odot$.  Because this is an emission line-selected sample, and care was taken to ensure robust photometry for all objects included,  after having performed the mass cuts, the resulting sample is complete in both SFR and mass in the given redshift ranges.

The dispersion in the star formation sequence of the HPS [O~II] emitters is about $\sigma_{\textrm{MS}} \sim 0.50$~dex, about 1.5 times as large as some other surveys \citep[e.g.,][]{noes07a, whit12}. Because our observations included just the [O~II] line, our SFRs do not take into account variations in abundance and ionization state.  This undoubtedly exacerbates the scatter in our star-forming main sequence by as much as 0.15 dex \cite[]{lop13}. Yet another source of additionally scatter is the sequence comes from invoking \cite{garn10} in the calculation of some $E(B-V)$, which can add up to $\sim0.28$ dex in scatter.  Finally, because our emission-line selected survey is sensitive to galaxies with lower SFRs, it is more prone to scatter in the star formation sequence, especially at lower masses. The larger ($\sigma_{\textrm{MS}} \sim 0.45$ dex) dispersion seen in the grism data of \cite{pirz13} supports this interpretation.  Additionally, \cite{bau13} used H$\alpha$ luminosities to determine the SFR of the $\sim73,000$ galaxies with redshifts of 0.05 $<$ $z$ $<$ 0.32 in the emission-line selected Galaxy And Mass Assembly (GAMA) survey, and found significant scatter in the main sequence, up to $\sim1$ dex in the highest mass bin.

The slope of this main sequence is somewhat shallower than comparable studies for the [O~II] emitting galaxies with redshift $z<0.5$.  A linear fit for the HPS sample produces
\begin{equation}
\textrm{log(SFR)}(\textrm{M}_{\sun}\:\textrm{yr}^{-1}) = (0.47 \pm 0.05)\: \textrm{log}(M_*/\textrm{M}_{\sun}) - (4.10 \pm 0.49)
\end{equation}
over the entire mass range in this sample. Noeske et al.\ (2007a) report a fit of 
\begin{equation}
\textrm{log(SFR)}(\textrm{M}_{\sun}\:\textrm{yr}^{-1}) = (0.67 \pm 0.08)\: \textrm{log}(M_*/\textrm{M}_{\sun}) - (6.19 \pm 0.78)
\end{equation}
for $M_*$ between $10^{10}$ and $10^{11}$ M$_\odot$ in the redshift range $0.2<z<0.7$. For the local universe, data from the Sloan Digital Sky Survey-Data Release 7 (SDSS-DR7) yield
\begin{equation}
\textrm{log(SFR)}(\textrm{M}_{\sun}\:\textrm{yr}^{-1}) = 0.55\: \textrm{log}(M_*/\textrm{M}_{\sun}) - 5.31
\end{equation}
with $\sigma = 0.349$ \cite[]{lop13}. Because the SDSS/GAMA survey includes galaxies with $z\leq0.1$ and the AEGIS sample \cite[]{noes07a} is in the range  $z = 0.2-0.7$, we would expect the HPS sample to fall in the same general category of slope and normalization. However, our galaxies have a slightly flatter slope; this could partly be due to the fact that HPS is an emission-line selected survey, meaning that the SFR-$M_*$ parameter space being explored is slightly different than the ones detected by continuum-selected surveys. 

\begin{figure}[!htbp]
\centering
\scalebox{0.55}
{\epsfig{file = 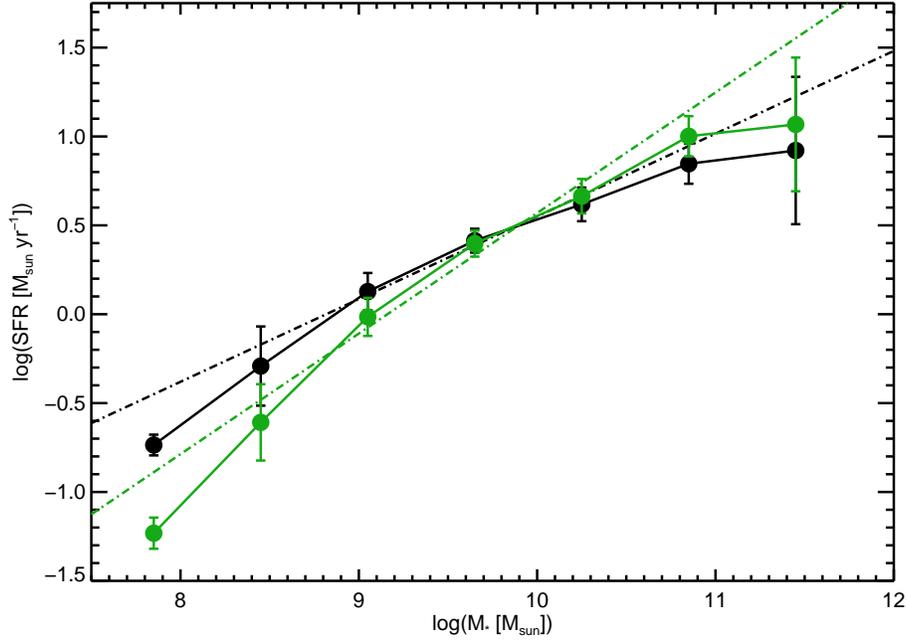}}
\caption{Median SFR vs. stellar mass for the HPS [O~II] emitters. The black circles represent the median SFRs in evenly-spaced mass bins, calculated without the metallicity correction to the [O~II] SFR indicator.  The green circles shows the median SFRs calculations that include a rough estimate of metallicity. The corresponding dash-dotted lines give the best linear fits to the data.  The error bars represent the uncertainty of the medians. Inclusion of the metallicity in the SFR calculation drives the sequence to a steeper slope. }
\label{fig:met}
\end{figure}

What is more likely affecting the slope of the main sequence, however, is that, as stated previously, the [O~II] star formation rate indicator is sensitive to metallicity, something we are not able to measure individually with the HPS data.  However, if we adopt a mass-metallicity relationship, we can estimate the effect that abundance shifts have on the slope. \cite{lop13} give a mass-metallicity relationship of
\begin{equation}
12 + \textrm{log(O/H)} = -10.8297 + 3.6478\: \textrm{log}(M_*/\textrm{M}_{\sun}) - 0.16706\: \textrm{log}(M_*/\textrm{M}_{\sun})^2
\end{equation}
This law, and the Kewley et al. (2004) relationship between SFR, [O~II] luminosity, and metallicity
\begin{equation}
\textrm{SFR([O~II], $Z$)}(\textrm{M}_{\sun}\:\textrm{yr}^{-1}) = \frac{7.9\times10^{-42} L(\textrm{[O~II]})\:(\textrm{ergs s}^{-1})}{(-1.75 \pm 0.25)[\textrm{12+log(O/H)}] + (16.73 \pm 2.23)}
\end{equation}
give the modified SFRs shown in Figure~\ref{fig:met}.  Including a rough estimate of the metallicity moves the higher mass galaxies to larger SFRs and the lower mass galaxies to lower SFRs, steepening the slope to
\begin{equation}
\textrm{log(SFR)}(\textrm{M}_{\sun}\:\textrm{yr}^{-1}) = (0.68 \pm 0.05)\: \textrm{log}(M_*/\textrm{M}_{\sun}) - (6.21 \pm 0.50)
\end{equation}
This is on par with what has been found in previous studies \citep[e.g.,][]{noes07a, whit12}, and illustrates the importance of metallicity to the [O~II] star formation rate indicator.

Figure~\ref{fig:SFRvM_PN} confirms the results of \cite{noes07a} and \cite{whit12} on the evolution of the main sequence, as the entire main sequence appears to shift downwards with redshift. The normalization of the main sequence decreases by almost a factor of two from the highest redshift bin (median $z\sim0.51$) to the lowest (median $z\sim0.1$).  

\begin{figure}[!htbp]
\centering
\scalebox{0.8}
{\epsfig{file = 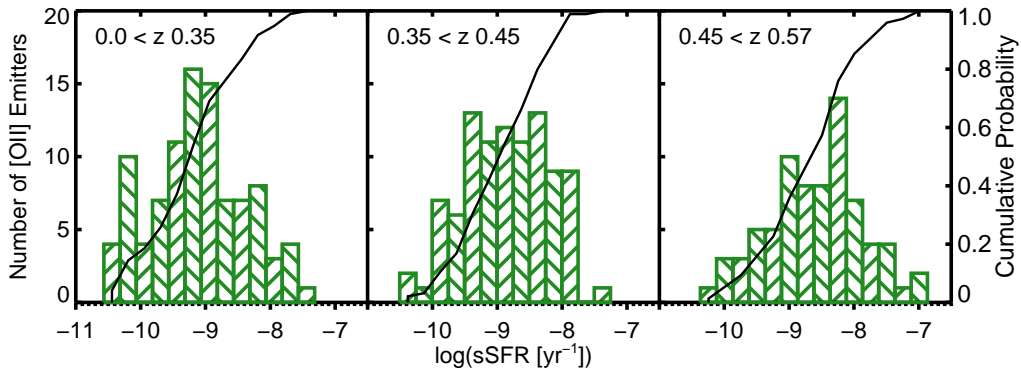}}
\caption{Distribution of the HPS [O~II] emitter sSFRs calculated via their [O~II] line luminosities. The black line is the cumulative probability distribution. See Table~\ref{tbl-2} for the median sSFRs in each redshift bin.}
\label{fig:SSFR}
\end{figure}

Another way to describe star formation in a galaxy is through its mass specific SFR (sSFR), which represents the time needed to build up the current stellar mass of the galaxy at its present day SFR \cite[]{pirz13}. The sSFR of galaxies has been shown at all redshifts to decrease as stellar mass increases \citep[e.g.,][]{bau05}. Figure~\ref{fig:SSFR} shows the distribution of the [O~II] emitters' sSFRs calculated via their [O~II] line luminosities. The median sSFRs per redshift bin are given in Table~\ref{tbl-2}. The distribution of the HPS [O~II] emitters shifts to larger sSFRs as redshift increases, indicating that galaxies at higher redshifts are more efficient at producing stars.  There is a trend for the more massive galaxies to be formed at higher redshifts and over a short burst timescales \cite{cow96}, but by $z\sim0.5$, their sSFR has decreased.  At that time, the lower mass galaxies have begun their (relatively) later onset of star formation, indicating that they have formed the bulk of their stars more recently but on longer evolutionary timescales.  These data are consistent with the concept of downsizing as described by \citeauthor{cow96}, where there is a smooth evolution downwards with redshift in the masses of star-forming galaxies.

\section{Equivalent Widths}
Figure~\ref{fig:EWvmass} plots the [O~II] rest-frame equivalent widths (EWs) for the HPS galaxies.  As found in other surveys \citep[e.g.][]{fum12} there is a weak anti-correlation between EW and stellar mass.  In the lowest mass bin, the median rest-frame equivalent width is three times greater than that in the highest mass bin.  For comparison, \citeauthor{fum12} found a factor of five shift in the EWs of H$\alpha$ between the mass bins of $10.0 < \log M/M_{\odot} < 10.5$, and $\log M/M_{\odot} > 11.0$.  The two measurements are not directly comparable, since, in addition to being in the redder range of the spectrum, the H$\alpha$ data were accumulated over a larger redshift range.   Nevertheless, the data do confirm the trend that emission-line equivalent widths in higher-mass galaxies are generally factors of
several lower than those in lower-mass objects.  Since the equivalent width of [O~II] is an indicator of the relative strength of star formation \citep{gil10, ciar13}, this trend is another reflection of the star-forming galaxies main sequence.

\section{Morphologies}
The HPS observations were conducted in regions of the sky with extensive ancillary data.  In particular, in the COSMOS and GOODS-N regions, deep imaging data is available from the {\sl Hubble Space Telescope\/} Advanced Camera for Surveys (ACS\null). These data allow us to characterize the morphology and size of the HPS [O~II] emitters via the Gini (G) and $M_{20}$ coefficients \citep{lotz04}.  To do this, we chose to use data in the F814W filter, as it is the only filter common to both the COSMOS \citep{scov07} and GOODS-N \citep{gia04} programs.

The $G$ coefficient represents the distribution of the flux values over the galaxy's pixels, and is thus similar to the concentration parameter $C$ \citep[e.g.][]{abr94, ber00, cons03}), which, loosely defined, is the ratio of light between inner and outer isophotes of a galaxy. However, $C$ is dependent on the spatial distribution of the light and is unable to differentiate between galaxies with off-centered light concentrations and those with shallow light profiles.  The Gini coefficient is sensitive to concentrations of bright pixels no matter where in the profile they occur, resulting in a more robust measure of concentration. 

The $M_{20}$ coefficient is defined as the second-order moment of the brightest 20\% of the flux of a galaxy. The second-order moment of flux ($M_{\textrm{tot}}$) is calculated by multiplying each pixel's flux by the squared distance to the center of the galaxy, and then summing over all the pixels:
\begin{equation}
M_{\textrm{tot}} = \sum_i^nM_i = \sum_i^nf_i[(x_i - x_c)^2 + (y_i-y_c)^2]
\end{equation}
where $f_i$ is the pixel flux and $x_c$, $y_c$ is the center of the galaxy \cite[]{lotz04}. Selecting the brightest 20\% of the pixels results in a parameter that is sensitive to the most important spatial distributions of the flux, such as multiple nuclei, bars, and star clusters.  Therefore, $M_{20}$ probes features indicative of star formation.

Gini and $M_{20}$ are dependent on the noise and resolution of the images, and are not reliable if the signal-to-noise per pixel drops below 2 \cite[]{lotz04}. Hence, galaxies with $\langle S/N \rangle$ less than this ($\sim15\%$ of the sample) were excluded from our analysis. Additionally, $G$ is only reliable to within $\sim15$\% at resolution scales better than 1000 pc while $M_{20}$ begins to degrade around 500 pc.  At the HPS redshift limit for [O~II] detection, the $\sim 0\farcs05$/pixel plate scale of ACS corresponds to a resolution of $\sim$ 330 pc. Assuming a spatial resolution no worse than $\sim 0\farcs1$, the highest redshift in our sample corresponds to a $\sim$ 650 pc resolution.  Therefore, while $M_{20}$ for the very highest redshift galaxies in the sample may have some degradation above 15\%,  $M_{20}$ and $G$ coefficients should still be for the most part reliable over the entire sample.

The morphology parameters of the COSMOS and GOODS-N [O~II] emitters were calculated using the \cite{lotz04} software, which uses SExtractor catalogues and segmentation maps. The distribution of these galaxies in $G$-$M_{20}$ space is plotted in Figure~\ref{fig:GvM20}, and the results are tabulated in Table~\ref{tbl-4}. \cite{lotz04} have shown that active or merging galaxies tend to lie above a certain threshold in $G-M_{20}$ parameter space. Only a fraction of the [O~II] emitters in this sample fall above this threshold.  Instead, the [O~II] emitters fall largely within the region of non-merging galaxies. Indeed, a cursory examination of $HST$ ACS images of the [O~II] emitters  confirms that most of these galaxies are not strongly distorted or visibly interacting.

This result is divergent from the results of \cite{pirz13}, who found that a significant fraction of the emission-line selected galaxies in their sample fell above the line delineating quiescent from active galaxies.  In particular, Figure 15 of \cite{pirz13} indicates that this effect is exaggerated for galaxies in which H$\alpha$ and [O III] lines were observed. It should be noted that \citeauthor{pirz13} measured their morphologies in the $\sim4350$~\AA~rest frame, while our measurements (and those of \cite{lotz04}) were performed using the F814W filter, i.e., the rest frame $R$-band.  Morphological parameters can change depending on the image wavelength \cite[]{tayl07}, but since both bands are redward of the 4000~\AA~break, this effect is not likely to be large.  Still, it is possible that the difference between the [O~II] galaxy samples are systematic.  Additionally, despite the fact that the galaxies found by the PEARS survey and HPS are both emission-line selected, at any given redshift, the HPS galaxies extend to much higher masses (see Figure~\ref{fig:SFRvM_PN}).  This may point to differing star-formation mechanisms at different masses, where higher mass galaxies are more likely to be undergoing normal star formation while lower mass galaxies are more likely to have disturbed morphologies.

While our [O~II] emitters do not appear to be interacting, it is possible that correlations between stellar mass and morphology exist.  Figure~\ref{fig:morph} shows $G$ and $M_{20}$ plotted as a function of both mass, SFR, and sSFR.  Calculation of the Pearson's correlation coefficient \cite[]{pear96} gives $r_{\textrm{SFR}, M_{20}} = -0.17$, $r_{\textrm{SFR}, G} = 0.016$, $r_{\textrm{M}, M_{20}} = -0.32$, $r_{\textrm{M}, G} = 0.098$, $r_{\textrm{sSFR}, M_{20}} = 0.18$, and $r_{\textrm{sSFR}, G} = -0.083$.  These indicate no significant correlations between these parameters. 

\section{Sizes}
In additional to quantifying their morphologies, we also measured the PSF-convolved physical sizes of our [O~II] emitters by performing photometry using a series of circular apertures. Following \cite{bond09}, we used these data to calculate each galaxy's half-light radius ($R_{50}$) in a manner that is more robust than simple isophotal measurements. The distribution of physical sizes is shown in Figure~\ref{fig:size_hist}. The [O~II] emitters from this survey are fairly small, with a median size of $R_{50} = 2.43^{4.20}_{1.39}$ kpc ($0.51^{0.87}_{0.27}$ arcseconds), where the upper and lower bounds are the $84^{\textrm{th}}$ and $16^{\textrm{th}}$ percentiles of the distribution, respectively.  There are two obvious outliers: a galaxy at a redshift of $z = 0.41$ with an angular size of $R_{50}$ = 4\farcs3, and another at $z=0.09$ with $R_{50}$ = 3\farcs8. These data have been excluded from our size analysis. Note that while the rest-frame bandpass of the $I$ filter changes with redshift, the range of redshifts considered here is fairly small and is unlikely to factor into the half-light radius results.

Figure~\ref{fig:sizevz} shows the sizes of these objects (in both kiloparsecs and arcseconds) as a function of redshift.  There is no significant evolution in the physical sizes of these galaxies in this redshift range, as demonstrated by the scatter in the upper panel of Figure~\ref{fig:sizevz}.  The fact that very few large galaxies are seen at lower redshifts is simply a volume effect, as the probability of seeing these rare objects is roughly proportional to $(1+z)^2$.  In the lower panel of Figure~\ref{fig:sizevz}, the dashed lines show, from top to bottom, the 84th, 50th, and 16th percentiles in the given redshift bins. These lines indicate that there is little evidence of evolution in the galaxy sizes in this sample. The median physical size varies from $R_{50}$ = 2.04 $\pm$ 0.29 kpc for redshifts $z<0.2$ to $R_{50}$ = 2.61 $\pm$ 0.56 kpc in the redshift bin $0.5<z<0.6$. The two numbers are therefore well within the uncertainties of the measurements.

\cite{pirz06} examined low-mass emission-line galaxies in the Hubble Ultra Deep Field and found similar scatter in the evolution of sizes with redshift, noting that over the redshift range corresponding to that of HPS, the angular sizes of the eGRAPES increased by a factor of $\sim2$, from $\sim$  0\farcs2 (1.33 kpc) at $z\sim0.7$ to $\sim$ 0\farcs5 (1.65 kpc) at $z\sim2$ in the 4350~\AA~rest frame.  While we measure the HPS [O~II] galaxy sizes using the $I$-band images (corresponding to $\sim R$-band rest frame), both data sets are redward of the 4000~\AA~break, so there should not be a significant difference between the two bands. Additionally, van der Wel et al.\ (2014) examined the sizes of galaxies in the CANDELS/3D-HST fields using the $HST$ WFC3 IR bands of  F125W, F140W, and F160W, just slightly redder than the $I$-band.  For the median mass of the HPS sample ($M  = 9.20 M_{\odot}$), they measure a median physical size of $R_{50} = 3.09  \pm 0.07$~kpc for $z > 0.5$ and $R_{50} = 2.69  \pm 0.07$~kpc for $0.5 < z < 1$, suggesting slow to moderate size evolution between these epochs. These measurements, combined with our results, are consistent with previous studies of disk-dominated late-type galaxies that indicate that there is little, if any, size evolution at redshifts $z < 1$ \citep{lil98, rav04, bar05}.

The size-mass distribution of the HPS [O~II] emitters is show in Figure~\ref{fig:massvsize}.  As has been shown in previous studies \citep[e.g.][]{shen03, van14}, there is an obvious correlation between the mass of galaxies and their size. The slope of the relation, $\alpha = \log R_{50} / \log M_* = 0.15 \pm 0.03$, is comparable to the $\alpha = 0.22 \pm 0.03$ measurement for late-type galaxies in the CANDELS/3D-HST field \citep{van14}, and midway between the values of $\alpha = 0.14$ and 0.39 found for low- ($M_* < 3.98 \times 10^{10} M_{\odot}$) and higher-mass late-type galaxies in SDSS \citep{shen03}.   Shen et al.\ (2003) showed that flatter slopes associated with the lower-mass galaxies are consistent with models in which feedback driven by galactic winds overcome the potential of the dark matter halos and suppress gas mass.   Assuming that the specific angular momentum of a galaxy is similar to that of its halo, less-massive late-type galaxies must have larger half-light radii, resulting in a flatter size-mass relation.

\section{Conclusions}
We presented an analysis of the physical properties of [O~II] emitting galaxies within the HETDEX pilot survey with $z < 0.57$.  The data quality of this sample of galaxies allowed us to achieve a substantial improvement in the study of properties of star-forming galaxies in this redshift range.  Stellar masses for these galaxies were determined by SED fitting.  The observed SFR vs. $M_*$ relation confirms the existence of the main sequence of galaxies, where SFRs decrease at lower masses and redshifts.  The sSFRs increase for lower stellar mass galaxies, supporting the idea that the galaxies that are more efficiently forming stars in this redshift range are of lower masses, while the larger galaxies are ending their star formation. Examination of the morphologies of these galaxies gives no indication that the [O~II] emitters are undergoing mergers, and there is no correlation between morphology parameters ($G$ and $M_{20}$) and mass, SFR, or sSFR. 

The HETDEX pilot survey has provided a test data set for the type of data produced in the main survey.  HETDEX will provide upwards of 10$^6$ [O~II] spectra, a data set that will serve to further refine the results presented here.  

We wish to thank the anonymous reviewer for their invaluable comments. We acknowledge N. Pirzkal for generously sharing his data, and J. Lotz for making her morphology code available and advising in its use. Additionally, we are grateful to J. Adams and G. Blanc for providing the photometry from HPS. We thank the Cynthia and George Mitchell Foundation for funding the Mitchell Spectrograph, known formerly as VIRUS-P. J.B is supported by the NSF Graduate Research Fellowship Program under grant DGE1255832.  This work was also funded by the NSF grant AST 09-26641. The Institute for Gravitation and the Cosmos is supported by the Eberly College of Science and the Office of the Senior Vice President for Research at The Pennsylvania State University.  Computational support and resources were provided by the Research Computer and Cyberinfrastructure Unit of Information Technology Services at The Pennsylvania State University. This research has made use of NASA's Astrophysics Data System Bibliographic Services.

\emph{Facility}: \facility{Smith (VIRUS-P)}

\bibliography{HPS_OII}

\begin{deluxetable}{cccccc}
\tabletypesize{\footnotesize}
\tablecaption{SED Fitting Photometric Bands\label{tbl-1}}
\tablecolumns{6}
\tablewidth{0pc}
\tablehead{\colhead{Field} & \colhead{Filter} & \colhead{Telescope} & \colhead{Instrument}  & \colhead{$\lambda_{c}$ (\AA)} & \colhead{$5\sigma$ Limit (AB)}}
\startdata
XMM-LSS & FUV & \emph{GALEX} & FUV Detector & 1528 &24.8\\
 & NUV &\emph{GALEX} & NUV Detector & 2271& 24.4\\
 & $u^*$ &CFHT &  MegaPrime/MegaCam & 3740 & 25.2 \\
 & $g'$  &CFHT & MegaPrime/MegaCam  &  4870 &  25.5\\
 & $r'$ &CFHT & MegaPrime/MegaCam & 6250 & 25.0\\
 & $i'$ &CFHT & MegaPrime/MegaCam & 7700 & 24.8\\
 & $z'$ &CFHT & MegaPrime/MegaCam  & 8900 & 23.9\\
 & Channel 1 & \emph{Spitzer} & IRAC& 37440 & 23.9\\
 & Channel 2 & \emph{Spizter} & IRAC & 44510 & 23.3\\
\tableline
MUNICS & FUV & \emph{GALEX} & FUV Detector  & 1528 &24.8\\
 & NUV &\emph{GALEX}  & NUV Detector & 2271& 24.4\\
 & $B$ &Calar Alto 3.5-m & LAICA & 4200 & 26.4 \\
 & $g'$ &Calar Alto 3.5-m  & LAICA & 4900 & $\sim$25.9\\
 & $i'$  &Calar Alto 3.5-m & LAICA  & 7700 & $\sim$24.3\\
 & $z'$ &Calar Alto 3.5-m & LAICA  & 9200 & $\sim$24.1\\
\tableline
COSMOS & FUV& \emph{GALEX}  & FUV Detector  & 1528 &24.8\\
 & NUV &\emph{GALEX} & NUV Detector & 2271& 24.4\\
 & $u^*$&CFHT  &  MegaPrime/MegaCam & 3740 & 26.5\\
 & $B$ &Subaru & Suprime-Cam & 4788 & 27.4\\
 & $V$ &Subaru & Suprime-Cam & 5730 & 27.2\\
 & $r'$ &Subaru & Suprime-Cam & 6600 & 26.9\\
 & $i'$&Subaru & Suprime-Cam  & 7850 & 26.9\\
 & $z'$ &Subaru & Suprime-Cam & 8700 & 25.6\\
 & $K$ &CFHT & WIRCam  & 21400 & 23.6\\
  & Channel 1 & \emph{Spitzer} & IRAC& 37440 & 23.9\\
 & Channel 2 & \emph{Spitzer} & IRAC & 44510 & 23.3\\
\tableline
GOODS-N  & FUV& \emph{GALEX} & FUV Detector & 1528 &24.8\\
 & NUV &\emph{GALEX} & NUV Detector & 2271& 24.4\\
 & $U$ &Mayall &  MOSAIC & 4065 & 27.1\\
 & $B$ &Suubaru & Suprime-Cam  & 4788 & 27.4\\
 & $V$ &Subaru & Suprime-Cam  & 5730 & 27.2\\
 & $r'$ &Subaru & Suprime-Cam & 6600 & 26.9\\
 & $i'$ &Subaru & Suprime-Cam & 7850 & 26.9\\
 & $z'$ &Subaru & Suprime-Cam  & 8700 & 25.6\\
 & $H+K'$&UH 2.2-m & QUIRC  & 20200 & 22.1\\
 & Channel 1 & \emph{Spitzer} & IRAC& 37440 & 23.9\\
 & Channel 2 & \emph{Spitzer} & IRAC & 44510 & 23.3\\
\enddata
\end{deluxetable}

\begin{figure}[!htbp]
\centering
\scalebox{0.6}
{\epsfig{file = 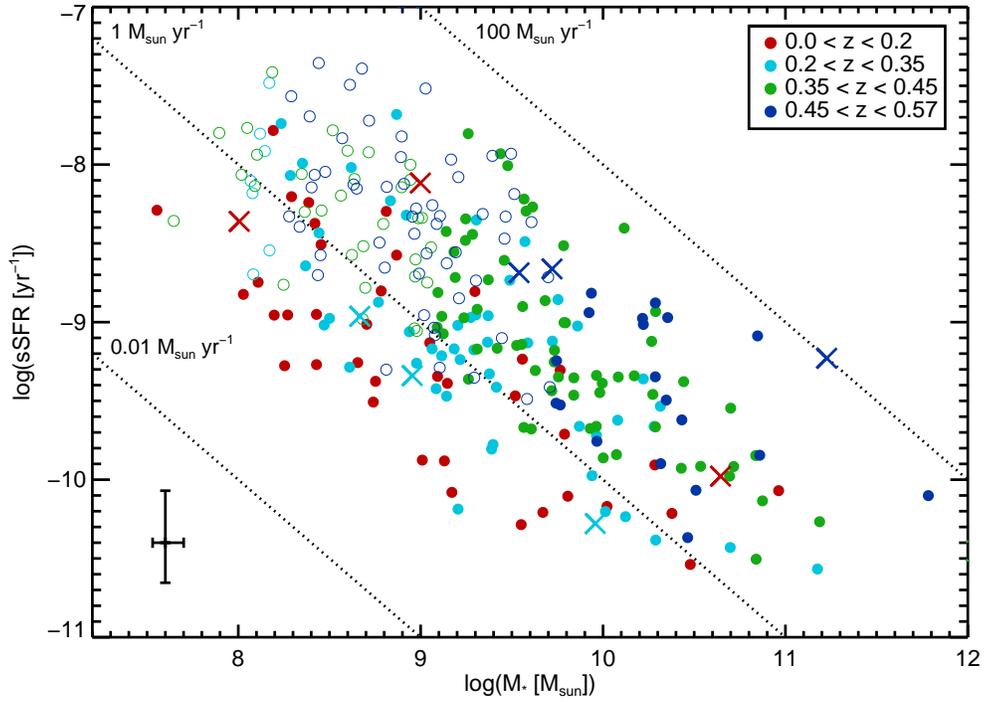}}
\caption{sSFR vs. $M_*$ for the HPS [O~II] emitters. The diagonal black dotted lines are loci of constant star formation with values 0.01, 1, and 100 M$_\odot$ yr$^{-1}$, from bottom to top. The crosses are possible AGN candidates due to their bright X-ray emission. Representative error bars are shown. The open circles represent galaxies below the stellar masses where SFR become incomplete in each redshift bin.}
\label{fig:SSFRvM}
\end{figure}

\begin{figure}[!htbp]
\centering
\scalebox{0.6}
{\epsfig{file = 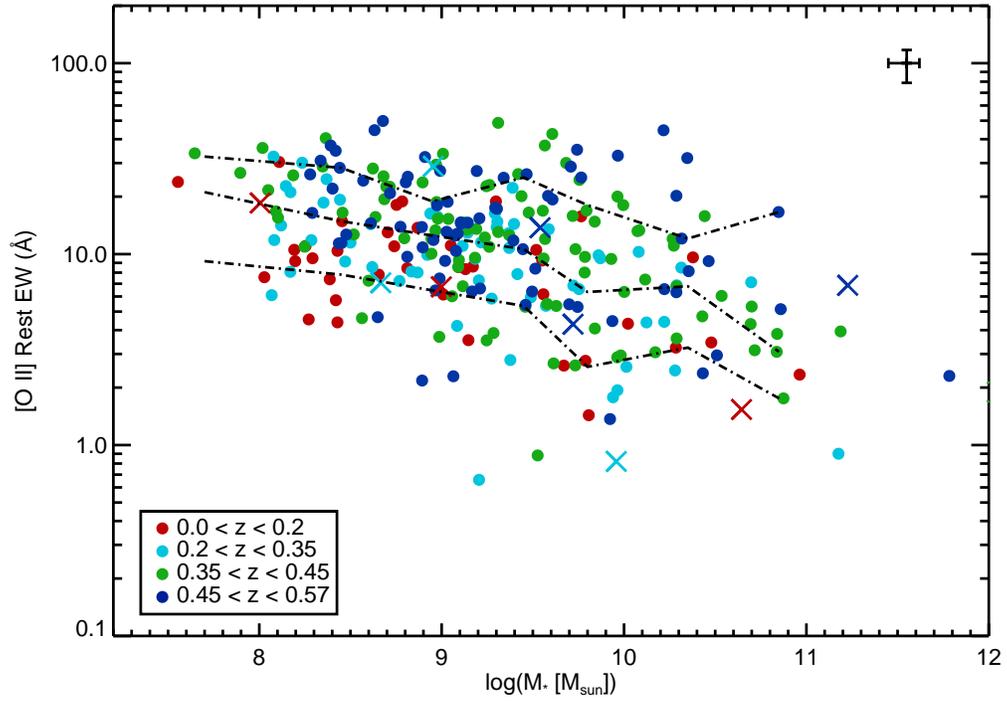}}
\caption{Rest equivalent widths of the HPS [O~II] lines as a function of stellar mass. The black dash-dotted lines are, from top to bottom, the 84th, 50th, and 16th percentiles in the given mass bins. The crosses are possible AGN candidates due to their bright X-ray emission. Representative error bars are shown.}
\label{fig:EWvmass}
\end{figure}

\begin{figure}[!htbp]
\centering
\scalebox{0.6}
{\epsfig{file = 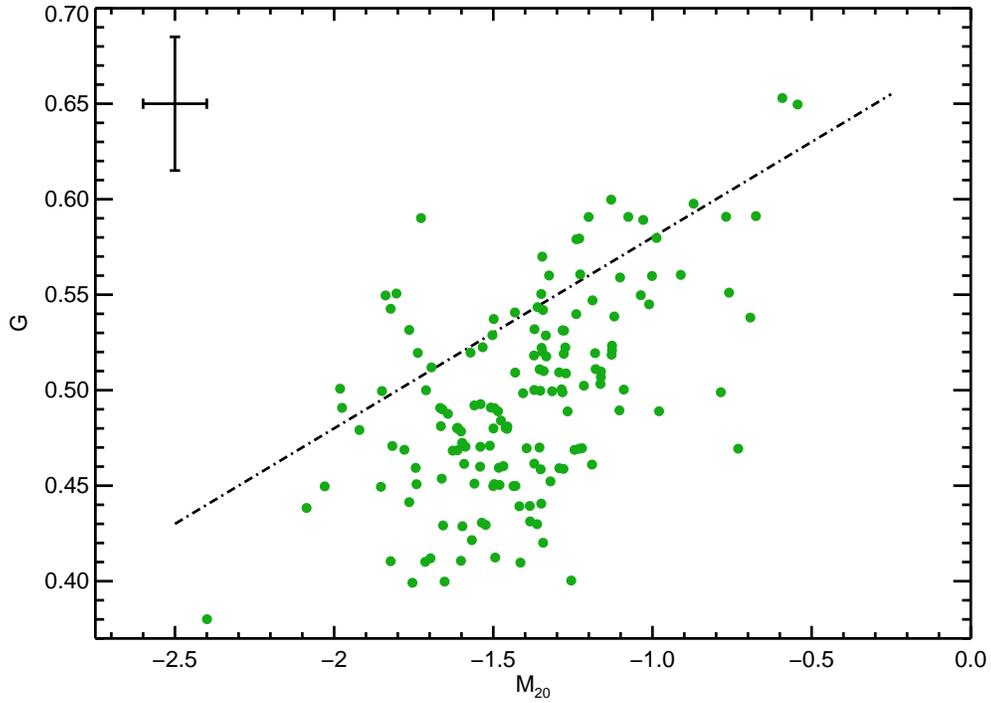}}
\caption{Morphological parameters $G$ and $M_{20}$ for the [O~II] emitters in the COSMOS and GOODS-N fields. The delineation between galaxies with disturbed morphologies and normal galaxies (black line) is taken from \cite{lotz04} and is for reference purposes only. Representative error bars are adapted from \cite{lotz06} and are typical for a $<$S/N$>$ = 2.5 galaxy.}
\label{fig:GvM20}
\end{figure}

\begin{figure}[!htbp]
\centering
\scalebox{0.6}
{\epsfig{file = 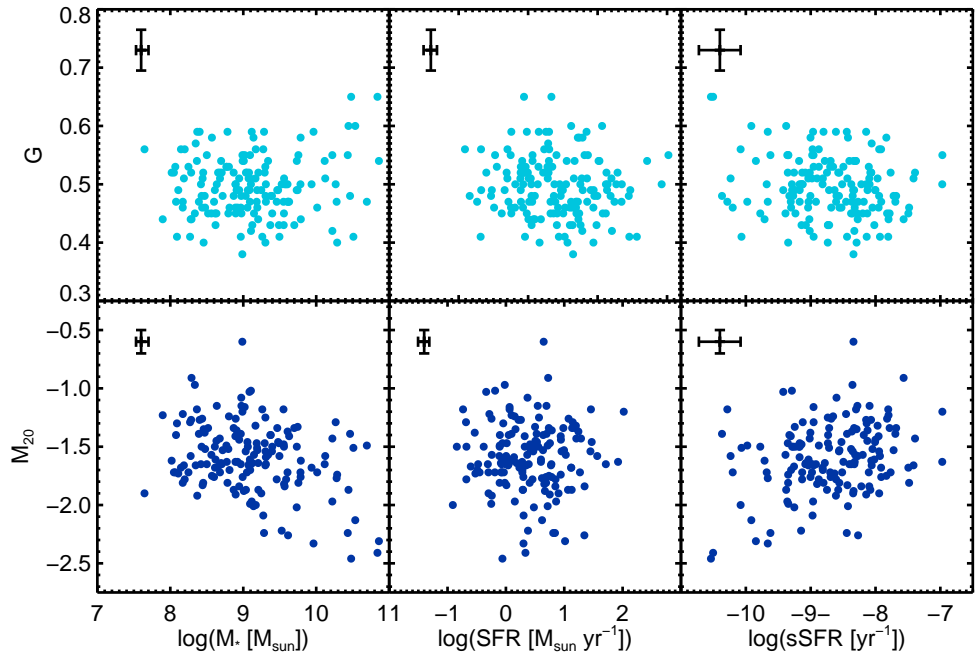}}
\caption{Morphological parameters $G$ and $M_{20}$ plotted as a function of stellar mass, SFR, and sSFR.  Pearson's correlation coefficient shows almost no significant correlation between the values. Representative error bars are shown, with those for $G$ and $M_{20}$ adapted from \cite{lotz06}, and are typical for a $<$S/N$>$ = 2.5 galaxy.}
\label{fig:morph}
\end{figure}

\begin{figure}[!htbp]
\centering
\scalebox{0.6}
{\epsfig{file = 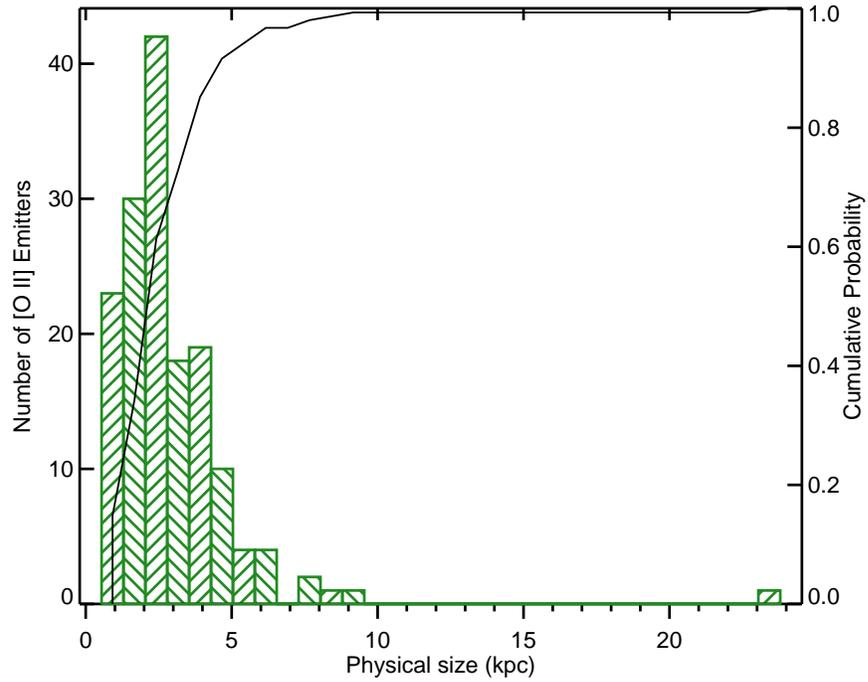}}
\caption{Physical half-light radii (R$_{50}$) of the HPS [O~II] emitters. The black line is the cumulative probability distribution. The median size is $R_{50} = 2.43^{4.20}_{1.39}$ kpc ($0.51^{0.87}_{0.27}$ arcseconds), where the upper and lower bounds are the $84^{\textrm{th}}$ and $16^{\textrm{th}}$ percentiles of the distribution, respectively.}
\label{fig:size_hist}
\end{figure}

\begin{figure}[!htbp]
\centering
\scalebox{0.6}
{\epsfig{file = 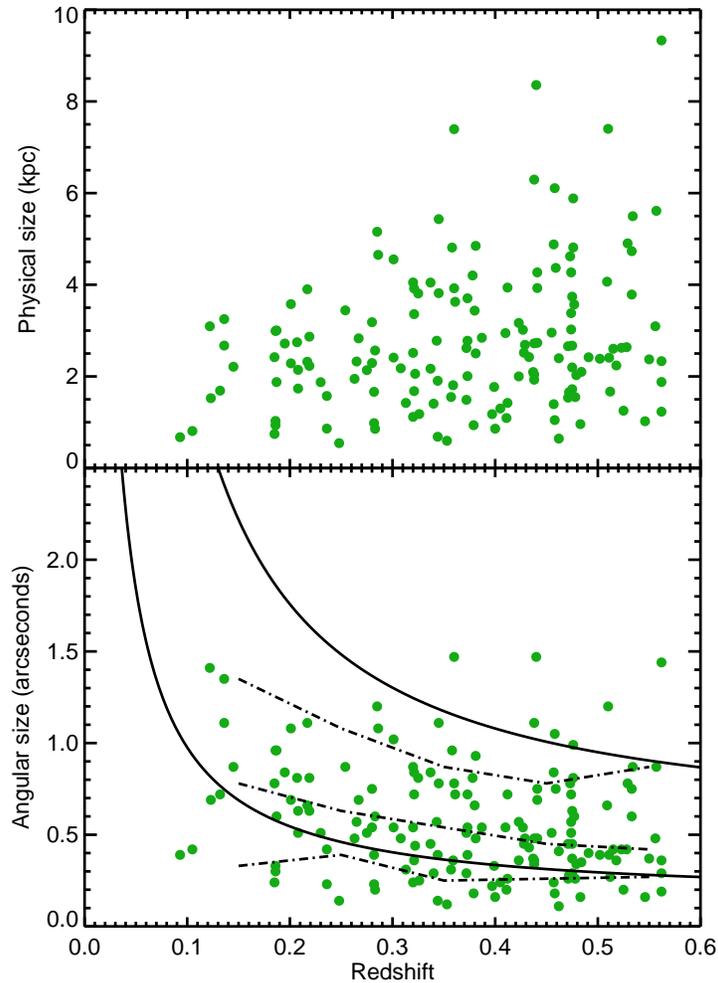}}
\caption{Physical sizes (R$_{50}$) of the HPS [O~II] emitters as a function of redshift. The solid black lines in the lower panel show what the angular size distribution would be assuming that the physical sizes of the galaxies do not evolve with redshift. The upper and lower lines represent the $84^{\textrm{th}}$ and $16^{\textrm{th}}$ percentiles of the size distribution at the highest sample redshift, corresponding to $R_{50}$=0\farcs27 and 0\farcs87 (5.8 and 1.8 kpc at $z\sim0.57$), respectively. The black dash-dotted lines are, from top to bottom, the 84th, 50th, and 16th percentiles in the given redshift bins. The median error on the physical sizes, adapted from \cite{lotz06}, are $\sim 0.1$ kpc, falling almost within the size of the plot symbols.}
\label{fig:sizevz}
\end{figure}

\begin{figure}[!htbp]
\centering
\scalebox{0.6}
{\epsfig{file = 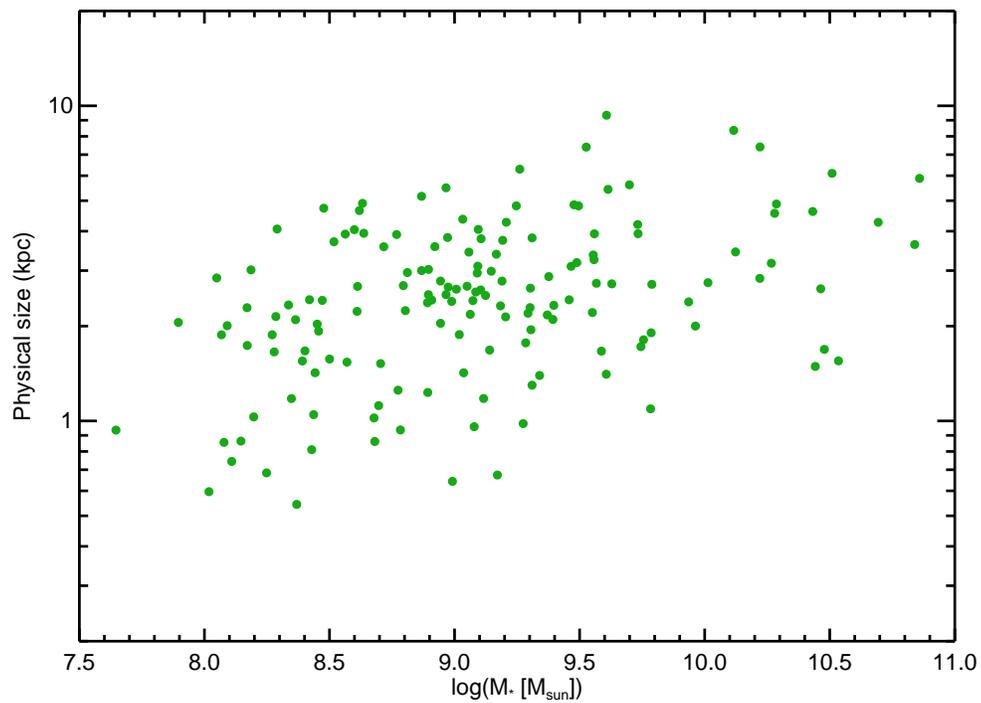}}
\caption{Physical sizes (R$_{50}$) of the HPS [O~II] emitters as a function of mass ($M_*$). The slope of the size-mass relation is $\alpha = 0.15$. The median error on the physical sizes, adapted from \cite{lotz06}, are $\sim 0.1$ kpc, falling almost within the size of the plot symbols.}
\label{fig:massvsize}
\end{figure}

\begin{deluxetable}{cccc}
\small
\tablecolumns{4}
\tablewidth{0pc}
\tablecaption{Median Logarithmic Galaxy Parameters by Redshift Bin\label{tbl-2}}
\tablehead{\colhead{Redshift} & \colhead{Mass ($M_*$/M$_\sun$)} & \colhead{SFR ($M_*$/M$_\sun$ yr$^{-1}$)} & \colhead{sSFR (yr$^{-1}$)}}
\startdata
$0<z<0.35$ & $9.13^{9.96}_{8.29}$  & $0.002^{0.60}_{-0.61}$ & $-9.13_{-9.91}^{-8.29}$\\
$0.35<z<0.45$ & $9.37^{10.17}_{8.64}$ & $0.53^{0.92}_{0.10}$ & $-8.9_{-9.55}^{-8.10}$\\
$0.45<z<0.57$ & $9.11^{10.22}_{8.63}$ & $0.70^{1.24}_{0.15}$ & $-8.56_{-9.35}^{-7.93}$\\
Total & $9.20^{10.08}_{8.45}$ & $0.38^{0.98}_{-0.11}$ & $-8.95_{-9.66}^{-8.10}$
\enddata
\tablecomments{The upper and lower bounds correspond to the $84^{\textrm{th}}$ and $16^{\textrm{th}}$ percentiles of the distributions.}
\end{deluxetable}

\clearpage

\begin{deluxetable}{ccccc}
\tablecolumns{5}
\tablewidth{0pt}
\tablecaption{Physical Parameters\label{tbl-3}}
\tablehead{\colhead{HPS ID} & \colhead{log($M_*$/M$_\sun$)} & \colhead{log(SFR) (M$_\sun$ yr$^{-1}$)}& \colhead{E(B-V)}& \colhead{$z$\tablenotemark{a}}}
\startdata
1 & 9.566 & -0.1026 & 0.037 & 0.400 \\
2 & 9.708 & 0.2954 & 0.210 & 0.462 \\
7 & 8.715 & 0.7933 & 0.243 & 0.385 \\
8 & 9.966 & 0.2098 & 0.003 & 0.562 \\
9 & 8.419 & 0.3534 & 0.204 & 0.466 \\
10 & 8.352 & 0.3602 & 0.224 & 0.29 \\
12 & 10.022 & -0.1480 & 0.279 & 0.18 \\
14 & 8.235 & 0.4956 & 0.216 & 0.305 \\
15 & 8.834 & 0.6050 & 0.300& 0.320 \\
16 & 9.142 & -0.3279 & 0.089 & 0.315 \\
19 & 10.847 & 1.7597 & 0.477 & 0.463 \\
20 & 9.143 & 0.5165 & 0.116 & 0.553 \\
21 & 9.583 & 0.0955 & 0.185 & 0.462 \\
23 & 9.389 & -0.4166 & 0.151 & 0.218 \\
24 & 8.082 & -0.6146 & 0.088 & 0.232 \\
26 & 9.438 & 1.5067 & 0.409 & 0.432 \\
29 & 9.298 & 0.4920 & 0.491 & 0.160 \\
31 & 8.955 & 0.6241 & 0.221 & 0.458 \\
32 & 9.923 & 0.9834 & 0.515 & 0.462 \\
33 & 9.742 & 0.2279 & 0.042 & 0.465 \\\enddata
\tablenotetext{a}{Redshifts as reported in \cite{ad11}}
\vspace{-0.1in}
\tablecomments{This table will be available in its entirety in a machine-readable form in the online journal. A sample is shown here as an indication of content.}
\end{deluxetable}

\clearpage

\begin{deluxetable}{cccc}
\tabletypesize{\footnotesize}
\tablecolumns{4}
\tablewidth{0pt}
\tablecaption{Morphological Parameters and Sizes\label{tbl-4}}
{\tablehead{\colhead{HPS ID} & \colhead{$G$} & \colhead{$M_{20}$} & \colhead{$R_{50}$ (kpc)}}}
\startdata
143 & 0.56 & -1.68 & 0.863 \\
146 & 0.49 & -1.5 & 2.179 \\
147 & 0.47 & -1.47 & 2.318 \\
149 & 0.5 & -1.72 & 1.407 \\
151 & 0.51 & -1.84 & 2.041 \\
152 & 0.55 & -1.97 & 2.834 \\
155 & 0.59 & -1.97 & 1.178 \\
158 & 0.56 & -2.0 & 0.674 \\
163 & 0.49 & -1.44 & 2.779 \\
165 & 0.49 & -1.5 & 2.227 \\
166 & 0.45 & -1.34 & 3.703 \\
167 & 0.58 & -1.76 & 1.905 \\
170 & 0.56 & -2.09 & 0.981 \\
171 & 0.47 & -1.4 & 0.855 \\
173 & 0.46 & -1.46 & 3.433 \\
175 & 0.5 & -1.78 & 1.665 \\
176 & 0.49 & -1.36 & 2.418 \\
177 & 0.43 & -1.64 & 2.502 \\
178 & 0.44 & -1.58 & 1.656 \\
179 & 0.53 & -1.72 & 0.684 \\
180 & 0.56 & -1.9 & 0.935 \\
185 & 0.56 & -1.78 & 1.03 \\
186 & 0.59 & -1.81 & 0.936 \\
188 & 0.49 & -2.02 & 4.2 \\
191 & 0.48 & -1.53 & 3.571 \\
192 & 0.51 & -1.83 & 1.422 \\
195 & 0.53 & -1.49 & 3.905 \\
198 & 0.45 & -1.5 & 2.327 \\
199 & 0.52 & -1.27 & 2.144 \\
200 & 0.5 & -1.63 & 1.231 \\
202 & 0.5 & -1.91 & 2.516 \\
204 & 0.47 & -2.26 & 5.431 \\
208 & 0.45 & -1.57 & 2.986 \\
209 & 0.52 & -1.87 & 3.0 \\
211 & 0.4 & -1.35 & 1.928 \\
212 & 0.52 & -1.81 & 2.029 \\
216 & 0.53 & -1.66 & 2.289 \\
217 & 0.41 & -1.18 & 6.294 \\
218 & 0.51 & -1.82 & 3.924 \\
224 & 0.5 & -1.68 & 2.386 \\
228 & 0.41 & -1.58 & 4.813 \\
230 & 0.51 & -1.66 & 3.016 \\
232 & 0.49 & -1.46 & 4.851 \\
235 & 0.65 & -2.46 & 1.687 \\
238 & 0.65 & -2.41 & 3.628 \\
243 & 0.46 & -1.65 & 2.516 \\
245 & 0.46 & -1.52 & 4.731 \\
246 & 0.48 & -1.38 & 2.413 \\
247 & 0.5 & -1.65 & 2.425 \\
248 & 0.6 & -1.87 & 1.49 \\
250 & 0.41 & -1.3 & 2.005 \\
252 & 0.54 & -1.65 & 8.358 \\
254 & 0.45 & -1.52 & 2.688 \\
255 & 0.55 & -1.65 & 1.572 \\
257 & 0.49 & -1.03 & 2.565 \\
264 & 0.49 & -1.9 & 4.815 \\
268 & 0.43 & -1.47 & 3.18 \\
270 & 0.53 & -1.63 & 1.252 \\
271 & 0.48 & -1.4 & 2.871 \\
272 & 0.58 & -1.75 & 1.537 \\
275 & 0.59 & -1.92 & 0.544 \\
276 & 0.41 & -1.29 & 1.877 \\
278 & 0.52 & -1.65 & 1.52 \\
282 & 0.43 & -1.48 & 3.362 \\
284 & 0.47 & -1.77 & 4.559 \\
289 & 0.5 & -1.59 & 1.298 \\
290 & 0.54 & -2.31 & 5.886 \\
291 & 0.55 & -1.81 & 1.809 \\
293 & 0.52 & -1.62 & 0.596 \\
294 & 0.54 & -1.88 & 2.619 \\
295 & 0.44 & -1.62 & 2.713 \\
297 & 0.48 & -1.54 & 1.947 \\
298 & 0.5 & -2.22 & 7.396 \\
299 & 0.47 & -1.22 & 1.736 \\
301 & 0.54 & -1.5 & 3.928 \\
302 & 0.52 & -1.87 & 1.422 \\
303 & 0.47 & -1.18 & 2.208 \\
305 & 0.47 & -1.65 & 3.809 \\
308 & 0.44 & -0.91 & 4.065 \\
317 & 0.47 & -1.39 & 1.55 \\
319 & 0.5 & -1.84 & 2.734 \\
320 & 0.55 & -1.2 & 2.396 \\
321 & 0.49 & -1.51 & 2.778 \\
322 & 0.6 & -2.13 & 1.551 \\
326 & 0.55 & -1.99 & 3.098 \\
328 & 0.45 & -1.15 & 2.659 \\
329 & 0.42 & -1.43 & 7.404 \\
330 & 0.49 & -1.73 & 0.744 \\
331 & 0.47 & -1.49 & 4.27 \\
332 & 0.43 & -1.61 & 4.652 \\
333 & 0.46 & -1.72 & 2.746 \\
335 & 0.48 & -1.34 & 5.162 \\
336 & 0.43 & -1.4 & 2.169 \\
337 & 0.54 & -1.18 & 2.095 \\
342 & 0.47 & -1.76 & 4.045 \\
343 & 0.55 & -2.24 & 4.619 \\
344 & 0.57 & -1.66 & 1.177 \\
346 & 0.48 & -1.08 & 3.819 \\
347 & 0.4 & -1.74 & 4.88 \\
348 & 0.59 & -2.24 & 1.769 \\
349 & 0.44 & -1.64 & 2.139 \\
350 & 0.41 & -1.51 & 6.107 \\
351 & 0.48 & -1.54 & 2.099 \\
352 & 0.45 & -0.97 & 2.333 \\
353 & 0.53 & -1.72 & 1.876 \\
354 & 0.43 & -1.34 & 5.613 \\
356 & 0.45 & -1.45 & 4.904 \\
358 & 0.47 & -1.41 & 2.723 \\
359 & 0.54 & -1.57 & 1.394 \\
361 & 0.4 & -1.25 & 2.638 \\
363 & 0.52 & -1.72 & 2.676 \\
368 & 0.48 & -1.54 & 4.367 \\
382 & 0.54 & -1.76 & 0.643 \\
388 & 0.47 & -1.78 & 2.423 \\
390 & 0.42 & -1.65 & 3.786 \\
393 & 0.46 & -1.41 & 2.373 \\
394 & 0.46 & -1.81 & 2.287 \\
396 & 0.49 & -1.34 & 3.094 \\
398 & 0.46 & -1.7 & 2.957 \\
399 & 0.45 & -1.57 & 3.917 \\
404 & 0.58 & -2.01 & 1.679 \\
406 & 0.41 & -1.39 & 5.493 \\
414 & 0.46 & -1.26 & 1.047 \\
416 & 0.46 & -1.53 & 3.742 \\
417 & 0.45 & -1.68 & 2.672 \\
418 & 0.49 & -1.33 & 1.721 \\
421 & 0.51 & -1.32 & 3.381 \\
422 & 0.45 & -1.26 & 3.026 \\
425 & 0.46 & -1.63 & 3.573 \\
427 & 0.53 & -1.24 & 2.239 \\
433 & 0.51 & -1.72 & 0.86 \\
435 & 0.51 & -1.58 & 3.438 \\
438 & 0.48 & -1.5 & 0.811 \\
439 & 0.52 & -1.43 & 1.022 \\
440 & 0.55 & -1.16 & 1.879 \\
441 & 0.52 & -1.87 & 4.271 \\
442 & 0.54 & -1.64 & 1.118 \\
443 & 0.51 & -1.7 & 4.053 \\
444 & 0.51 & -1.66 & 2.409 \\
450 & 0.52 & -1.67 & 2.604 \\
451 & 0.5 & -1.02 & 2.001 \\
454 & 0.59 & -2.33 & 1.668 \\
458 & 0.38 & -0.6 & 3.249 \\
459 & 0.59 & -1.27 & 2.055 \\
465 & 0.45 & -1.5 & 2.843 \\
469 & 0.44 & -1.23 & 3.166 \\
470 & 0.52 & -1.72 & 9.336 \\
471 & 0.5 & -1.29 & 1.092 \\
472 & 0.47 & -1.37 & 0.959 \\
473 & 0.5 & -1.72 & 2.197 \\
475 & 0.52 & -1.65 & 2.947 \\
476 & 0.52 & -1.47 & 3.941 \\
479 & 0.47 & -1.61 & 2.626 \\
\enddata
\end{deluxetable}

\end{document}